\documentclass{elsart3p}

\journal{J. Alloys Compd.}
\date{30 June 2006}

\usepackage{graphicx}
\usepackage[T1]{fontenc}
\usepackage{amssymb}
\usepackage{amsmath}

\begin{document}

\begin{frontmatter}

\title{On the solubility of Nd$^{3+}$ in Y$_3$Al$_5$O$_{12}$}

\author[1]{D. Klimm\corauthref{cor1}}\ead{klimm@ikz-berlin.de}, \corauth[cor1]{corresponding author} \author[1]{S. Ganschow}\ead{ganschow@ikz-berlin.de}, \author[2]{A. Paj\k{a}czkowska}\ead{apajaczkowska@gmail.com},\and
\author[2]{L. Lipi\'nska}\ead{Ludwika.Lipinska@itme.edu.pl}

\address[1]{Institute of Crystal Growth, Max-Born-Str. 2, 12489 Berlin, Germany}
\address[2]{Institute of Electronic Materials Technology, 01-919 Warsaw, Poland}

\begin{abstract}
Neodymium doped yttrium aluminum garnet powders (Nd:YAG or (Y$_{1-x}$Nd$_x$)$_3$Al$_5$O$_{12}$, $x=0.15$, 0.25, or 0.30, respectively) were prepared by a sol-gel technique. By DTA measurements up to $2000^{\:\circ}$C eutectic and liquidus temperatures could be determined. Exothermal peaks in the second and subsequent DTA heating runs indicate that the crystallized DTA samples are not in equilibrium. The section Y$_3$Al$_5$O$_{12}$--Nd$_3$Al$_5$O$_{12}$ of the concentration triangle Al$_2$O$_3$--Nd$_2$O$_3$--Y$_2$O$_3$ is proposed on the basis of thermodynamic calculations that allows to explain the experimental results by the balance of metastable phase states in the previously crystallized DTA samples.
\end{abstract}


\end{frontmatter}

\section{Introduction}

\begin{figure}[htb]
\includegraphics[width=0.46\textwidth]{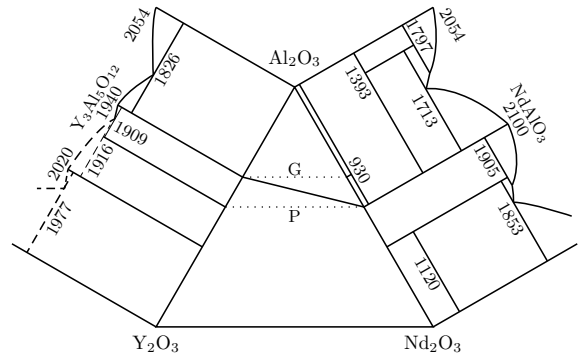}
\caption{Ternary concentration triangle with ternary sections G (garnets) and P (perovskites). The solid ternary line limits the partial system Al$_2$O$_3$--Y$_3$Al$_5$O$_{12}$--NdAlO$_3$.}
\label{Fig:ternary}
\end{figure}

Neodymium doped yttrium aluminum garnet (Nd:YAG or Y$_{1-x}$Nd$_x$)$_3$Al$_5$O$_{12}$) is one of the most important laser materials. Crystals are available from different commercial suppliers with dopant concentrations up to 2.5\,at\% ($x\leq0.025$) \cite{Fee06}. Bakradze et al. \cite{Bakradze68} (see also reference in \cite{ACerS301}, Fig. 04602) studied the subsolidus of the ternary system Al$_2$O$_3$--Nd$_2$O$_3$--Y$_2$O$_3$ and found as maximum solubility for Nd $\approx12$~wt-\% in the garnet phase, corresponding to (Y$_{0.82}$Nd$_{0.18}$)$_3$Al$_5$O$_{12}$ ($x=0.18$). The Nd:YAG samples were obtained by cooling from PbO/PbF$_2$ solutions held at $1200^{\:\circ}$C. Maximum solubility was observed along the ``garnet section'' ``G'' of the concentration triangle Fig. \ref{Fig:ternary}. 

Recently, some of the present authors reported on Nd:YAG powders that were prepared by a sol-gel technique followed by an annealing step at temperatures down to $T=800^{\:\circ}$C and found the maximum doping concentration (Y$_{0.725}$Nd$_{0.275}$)$_3$Al$_5$O$_{12}$ ($x=0.275$) \cite{Lipinska06}. The results were confirmed by X-ray and chemical (ICP-AES and SEM-EDX) analysis. If single crystals were melt-grown from such sol-gel powders by the micro-pulling-down technique, already $x\geq0.08$ led to the formation of a second (perovskite) phase as the solubility limit of Nd$^{3+}$ in Y$_3$Al$_5$O$_{12}$ was exceeded.

Both reports \cite{Bakradze68,Lipinska06} indicate, that the solubility limit of Nd in YAG may be large in the subsolidus but drops drastically approaching the liquidus $T\gtrsim1900^{\:\circ}$C. Unfortunately, the experimental determination of equilibrium phase relations in the subsolidus region is difficult, as the time that is needed to reach equilibrium rises exponentially if $T$ is lowered.

The present study will report on DTA measurements with the same sol-gel samples that were already used in the previous study \cite{Lipinska06}. The results allow to propose the topology of the garnet section ``G'' Y$_3$Al$_5$O$_{12}$--Nd$_3$Al$_5$O$_{12}$ of the ternary phase diagram Fig. \ref{Fig:ternary}.

\section{Experimental}

The preparation of nanocrystalline Nd:YAG powders (Y$_{1-x}$Nd$_x$)$_3$Al$_5$O$_{12}$ with $x=0.15, 0.25, 0.30$ was performed in aqueous solution starting from Y$_2$O$_3$ and Nd$_2$O$_3$ that were dissolved in acetic acid and from aluminum nitrate and is described in more detail elsewhere \cite{Lipinska06}. $12-18$~mg of the apparently white Nd:YAG powders were filled in NETZSCH standard tungsten DTA crucibles. (The mass of the empty crucibles is $500-600$~mg.) A STA 409C thermal analyzer with graphite furnace and tungsten sample holder (W/Re thermocouples) was used to perform the DTA/TG measurements in flowing Ar (40~ml/min). The following $T$ program was performed twice for each sample: 1) From room temperature to $2000^{\:\circ}$C with 15~K/min. 2) Down to $500^{\:\circ}$C with 20~K/min. 3) Up to $2000^{\:\circ}$C with 15~K/min. 4) Down to $500^{\:\circ}$C with 20~K/min. This means, 4 heating/cooling runs were performed in total for each sample. It must be noted, that the signal/noise ratio of DTA signals in the region of very high $T\lesssim2000^{\:\circ}$C is large as compared to lower $T$: 1) A large portion of heat is exchanged by radiation and not by conduction through the sample holder. 2) The mass ratio sample/crucible is worse than for DTA crucibles that are used for lower $T$, e.g. from platinum.

\begin{figure}[htb]
\includegraphics[width=0.42\textwidth]{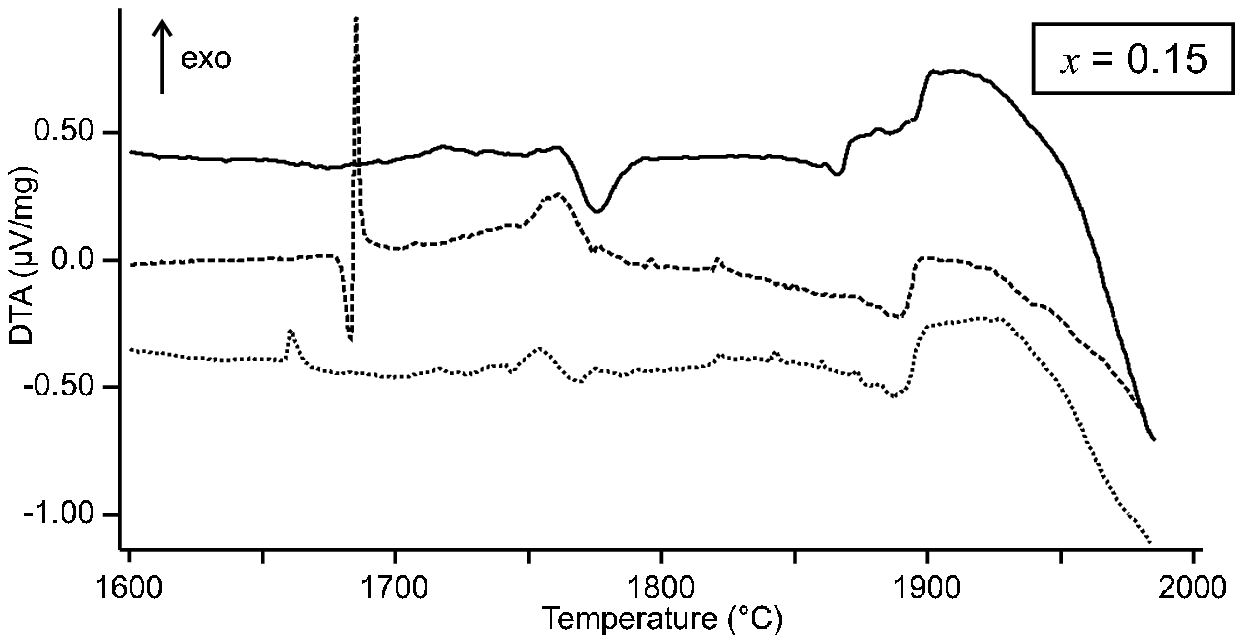}
\includegraphics[width=0.42\textwidth]{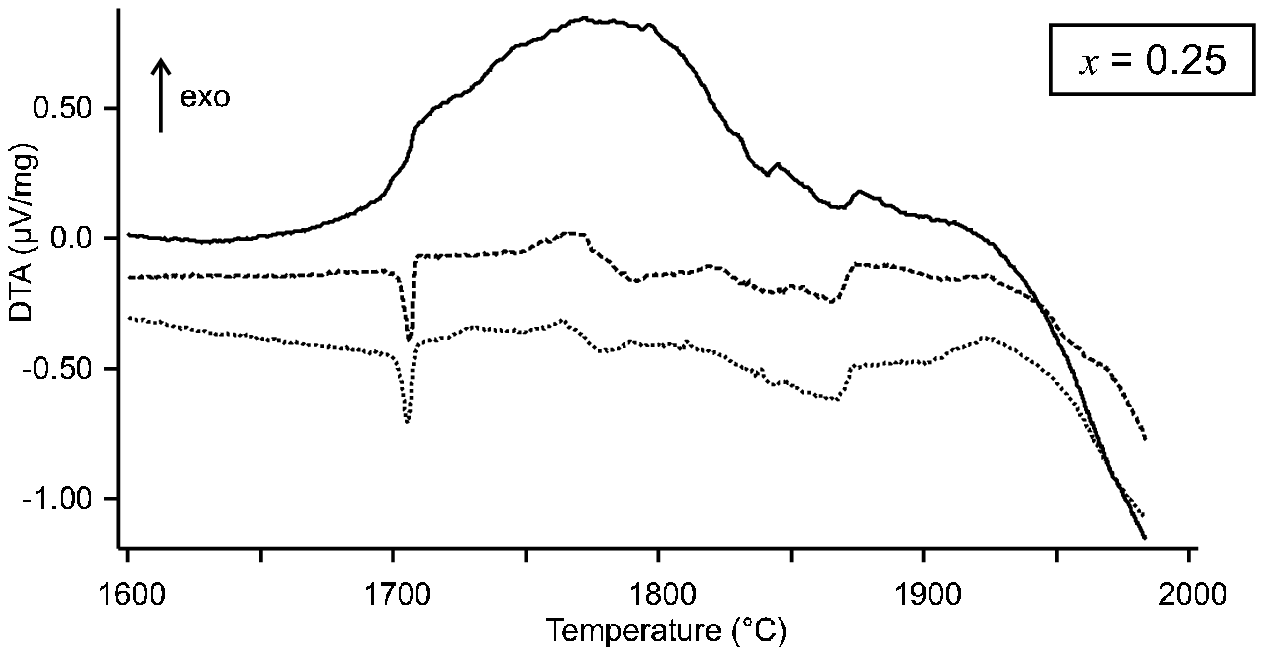}
\includegraphics[width=0.42\textwidth]{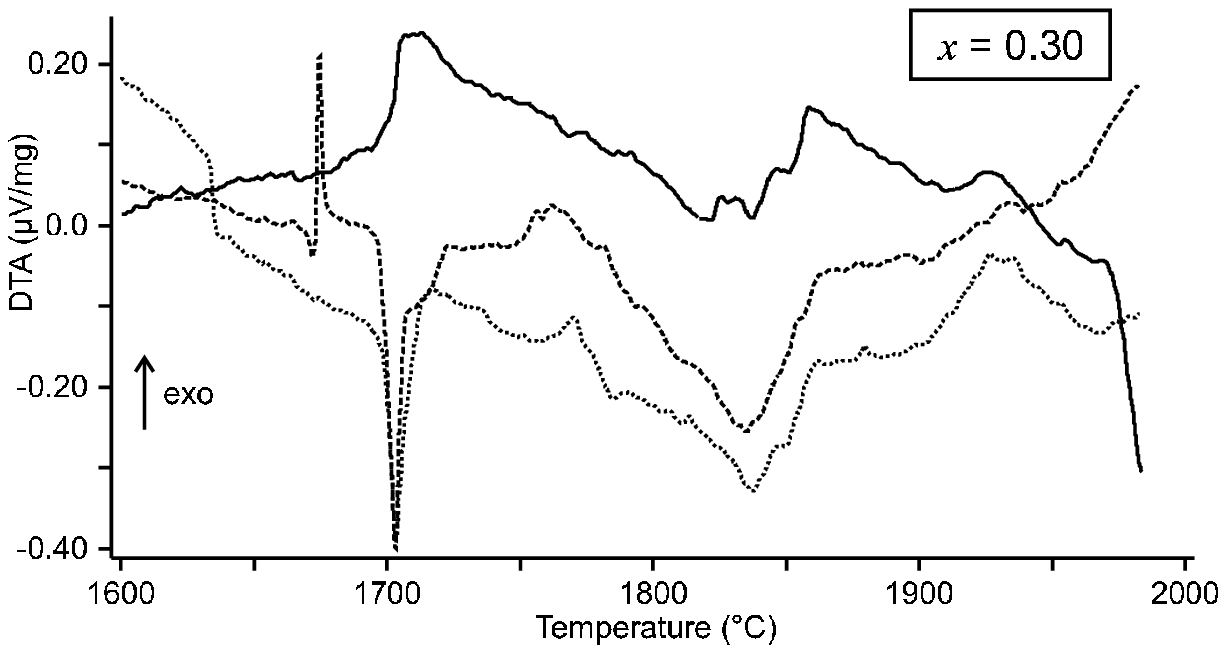}
\caption{1$^\text{st}$ (solid), 2$^\text{nd}$ (dashed), and 3$^\text{rd}$ (dotted) DTA heating run of
the Nd:YAG samples. (Nd concentration $x$ denoted in the respective graphs.)}
\label{fig:DTA}
\end{figure}

No significant mass loss was found during the DTA/TG measurements, indicating that the Nd:YAG powders were free of volatiles. After the measurements the samples formed solidified melts covering the bottom of the W crucibles. The purple color of this melt deepened considerably with doping level. The DTA curves from the first heating runs of all samples were obviously different from the curves of the heating runs 2 -- 4. The cooling curves showed strong exothermal peaks at different $T$, indicating undercooling upon crystallization and are therefore not well suited for the construction of an equilibrium phase diagram. Fig. \ref{fig:DTA} shows the DTA heating curves 1 -- 3 of the samples.

\section{Discussion of DTA curves}

A solid that is heated above the fusion point $T_\text{f}$ melts under consumption of the heat of fusion $\Delta H_\text{f}$. Accordingly, melting is an endothermal process. If the melting phase transformation solid $\rightarrow$ liquid proceeds through a 2-phase region of the phase diagram, the solid melts incongruently. This may be the case for mixed crystals (solid solutions, melting starts at solidus $T_\text{sol}$ and terminates at liquidus $T_\text{liq}$) as well as for arbitrary compositions within an eutectic system (melting starts at eutectic $T_\text{eut}$ and terminates at $T_\text{liq}$). All melting events are endothermal.

The 15\% Nd:YAG sample shows such normal behavior (solid line Fig. \ref{fig:DTA} top): The first peak with onset $T_\text{sol}=1764^{\:\circ}$C can be attributed to the solidus. It is followed by a broad shoulder. The return to the baseline indicates termination of the melting at the liquidus $T_\text{liq}=1898^{\:\circ}$C. Contrary, the 25\% and 30\% samples do not show any endothermal peaks that can be related to the onset of a melting process. Instead, broad exothermal effects starting around $1700^{\:\circ}$C can be observed for both samples.

The 25\% Nd:YAG sample shows in the 2$^\text{nd}$ and all following runs a sharp endothermal effect probably related to a eutectic with onset at $T_\text{eut}^{(25\%)}=1703^{\:\circ}$C. For the 30\% sample one finds $T_\text{eut}^{(30\%)}=1700^{\:\circ}$C. Such eutectic melting is never observed in the 15\% sample. This behaviour shows, that the composition (Y$_{0.85}$Nd$_{0.15}$)$_3$Al$_5$O$_{12}$ is still within the solubility range of Nd$_3$Al$_5$O$_{12}$ in Y$_3$Al$_5$O$_{12}$ and the other compositions are beyond the solubility limit. Between $T_\text{sol}$ of 15\% Nd:YAG and $T_\text{eut}$ of the higher doped samples some endo- and exothermal effects occur until the DTA curve returns to the baseline. This return (extrapolated offset) marks the liquidus that is clearly for the lower doped samples $T_\text{liq}^{(15\%)}=1897^{\:\circ}$C or $T_\text{liq}^{(25\%)}=1873^{\:\circ}$C, respectively. For (Y$_{0.70}$Nd$_{0.30}$)$_3$Al$_5$O$_{12}$ $T_\text{liq}^{(30\%)}=1860^{\:\circ}$C was determined by repeated measurements. (Not all DTA curves are shown in Fig. \ref{fig:DTA}.)

It is surprising that for all compositions the 1$^\text{st}$ heating curve is greatly different from the following ones. Moreover, beginning endothermal peaks do sometimes knock over in sharp exothermal effects (e.g. 2$^\text{nd}$ heating of the 15\% sample near $1680^{\:\circ}$C). Such endo-/exothermal effects during heating show bad reproducibility and cannot be explained by equilibrium processes. Instead, subsolidus transformations must be responsible for exothermal and non-reproducible effects, as in the subsolidus range the transformation between phases is governed by slow diffusion processes and tends thus to proceed beyond equilibrium. The following section will explain these phenomena on the thermodynamic basis.

\section{Discussion of phase equilibria}

The (Y$_{1-x}$Nd$_x$)$_3$Al$_5$O$_{12}$ compositions that were investigated in this study can be found on line ``G'' of Fig. \ref{Fig:ternary}. This line starts left at Y$_3$Al$_5$O$_{12}$ that is a congruently melting quasi-binary compound of the rim system Al$_2$O$_3$--Y$_2$O$_3$. In the opposite rim system Al$_2$O$_3$--Nd$_2$O$_3$ the perovskite NdAlO$_3$ is the only congruently melting compound. Experimental phase diagrams of this system can be found in \cite{Toropov61,Mizuno79,Coutures85} (see also compilation in \cite{ACerS301}, Figs. 2342, 6439, and 9262), for $T\gtrsim1300^{\:\circ}$C and show in addition to NdAlO$_3$ the $\beta$-alumina type phase Al$_{11}$NdO$_{18}$ and the monoclinic (space symmetry group $P\;2_1 /c$) Nd$_4$Al$_2$O$_9$ that melt both incongruently.

The ``neodymium aluminum garnet'' Nd$_3$Al$_5$O$_{12}$ was, to the authors' know\-ledge, not yet prepared. Attempts to produce the phase by the same sol-gel technique like the mixed garnets (Y$_{1-x}$Nd$_x$)$_3$Al$_5$O$_{12}$, or by crystallization from PbO/PbF$_2$ solutions, failed. Nd$_3$Al$_5$O$_{12}$ is shown as a notional end member in the concentration triangle Al$_2$O$_3$--Nd$_2$O$_3$--Y$_2$O$_3$ \cite{Bakradze68}. Wu and Pelton \cite{Wu92} carried out a critical assessment of 15 different RE$_2$O$_3$--Al$_2$O$_3$ (RE -- rare earth metal) systems, among them RE = Nd (but not Y) and calculated thermodynamic properties for Nd$_3$Al$_5$O$_{12}$ and refined excess enthalpies for the binary melts on the basis of a quasi-chemical model. These data are incorporated in the data bases coming with FactSage \cite{FactSage5_4_1} and were used for the calculation of the Al$_2$O$_3$--Nd$_2$O$_3$ rim system in Fig. \ref{Fig:ternary}. Following this calculation, Nd$_3$Al$_5$O$_{12}$ should be stable only for $T<930^{\:\circ}$C and decomposes at higher $T$ to NdAlO$_3$ and $\alpha$-Al$_2$O$_3$. As the upper stability limit of Nd$_3$Al$_5$O$_{12}$ is so low one can expect that the preparation of this compound will be very difficult, if not practically impossible.

Unfortunately, the very thorough analysis by Wu and Pelton \cite{Wu92} did not consider the system Y$_2$O$_3$--Al$_2$O$_3$ and well assessed thermodynamic data $G(T)$ for Y$_3$Al$_5$O$_{12}$ are not found in the FactSage database. However, the properties of RE-Al garnets are known to be very similar, if the ionic radii $r_\text{RE}$ are similar too. In octahedral environment one finds $r_{\text{Y}^{3+}}=104.0$~pm. The most similar RE is holmium with $r_{\text{Ho}^{3+}}=104.1$~pm. The ``mixer'' function of FactSage that is based on a Born-Haber cycle allows to calculate data for Y$_3$Al$_5$O$_{12}$ from the notional reaction

\begin{equation}
	2~\underbrace{\text{Ho}_3 \text{Al}_5 \text{O}_{12}}_{(1)} +~3~\underbrace{\text{Y}_2 \text{O}_3}_{(2)} \rightarrow 2~\underbrace{\text{Y}_3 \text{Al}_5 \text{O}_{12}}_{(3)} +~3~\underbrace{\text{Ho}_2 \text{O}_3}_{(4)}
	\label{eq:reaction}
\end{equation}

where all data except for (3) are known. With (\ref{eq:reaction}) one obtains for Y$_3$Al$_5$O$_{12}$

\begin{eqnarray}
	G(T) &=& -7387342.19 + 3854.78457\;T \nonumber \\ 
	&& - 3.7656\times10^{-3}\;T^2 + 6326731.0/T \nonumber \\
	&& -8283.84604\;\sqrt{T} -170451139.0/T^2 \nonumber \\
	&& - 573.3126\;T\ln T \label{eq:G} \\
	c_p(T) &=& 573.3126 + 7.5312\times10^{-3}\; T \nonumber \\ 
	&& -12653462.0/T^2 -2070.9615/\sqrt{T} \nonumber \\
	&& + 1.02270684\times10^9/T^3 \label{eq:c_p}
\end{eqnarray}

for $298\leq T/\text{K}\leq2213$ ($G$ given in J/mol, $c_p$ in J/(mol K), $\Delta H(298)=-7250905.48$~J/mol, $S(298)=285.461831$~J/(mol K)).

\begin{figure}[htb]
\includegraphics[width=0.42\textwidth]{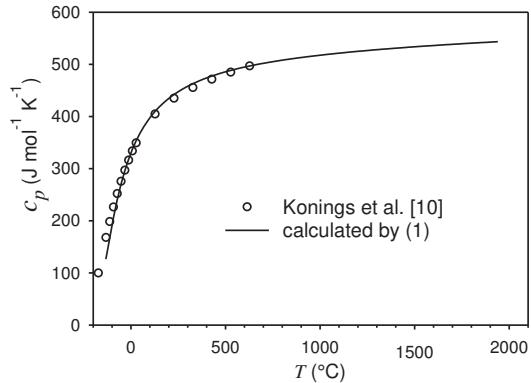}
\caption{Specific heat capacity of Y$_3$Al$_5$O$_{12}$ calculated from (\ref{eq:reaction}) together with experimental data \cite{Konings98}.}
\label{fig:cp_YAG}
\end{figure}

Fig. \ref{fig:cp_YAG} compares the results of the ``mixer'' calculation (\ref{eq:c_p}) with experimental data that were obtained by Konings et al. \cite{Konings98} using adiabatic calorimetry for $T=5-420$~K and by drop calorimetry for $T=470-880$~K with an experimental error of $<0.35$\%. The theoretical data from (\ref{eq:reaction}) approach the experimental data around room temperature with a difference of 1\%. The difference drops with $T$ and reaches 0.05\% at the experimental limit.

The calculated thermodynamic data $G(T)$ for Y$_3$Al$_5$O$_{12}$ (\ref{eq:G}) result at the melting point $1940^{\:\circ}$C \cite{Bondar84,Mah92} in a heat of fusion $\Delta H_\text{f}=300.9968$~kJ/mol. (Xiao, Derby \cite{Xiao94} use for Y$_3$Al$_5$O$_{12}$ $\Delta H_\text{f}=270.3$~kJ/mol without further reference.) Together with the data for Nd$_3$Al$_5$O$_{12}$ from FactSage \cite{FactSage5_4_1} one can calculate the ``G'' section through the concentration triangle Fig. \ref{Fig:ternary}, if reasonable estimations are done for the 2 mixed phases liquid (\textit{l}) and garnet (\textit{g}). As excess enthalpies $G_\text{ex}$ are unknown for both phases, ideal mixing and $G=G_0+G_\text{ideal}$ ($G_0$ -- weighed $G$ for the pure components of the mixed phase) was assumed. The influence of neglecting $G_\text{ex}$ was checked for different other RE$_2$O$_3$--Al$_2$O$_3$ systems where experimental phase diagrams as well as $G_0$ and $G_\text{ex}$ are known \cite{ACerS301,FactSage5_4_1}. It was found, that the deviations for liquidus temperatures are typically up to $T_\text{liq}\lesssim150$~K; but the topology of the whole phase diagram (occurrence of eutectics and of 1-, 2- or 3-phase rooms) was never changed.

\begin{figure}[htb]
\includegraphics[width=0.46\textwidth]{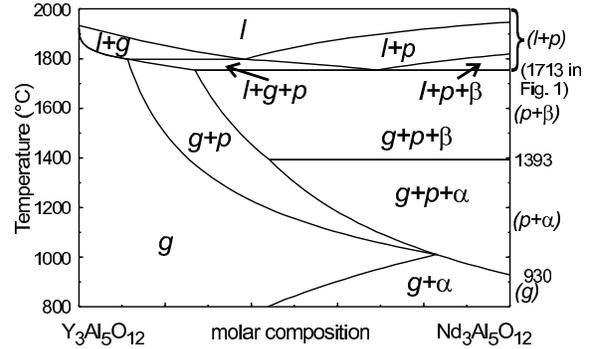}
\caption{Calculated section ``G'' through the ternary phase diagram Fig. \ref{Fig:ternary}: \textit{l} -- ideal liquid, \textit{g} -- ideal garnet, \textit{p} -- perovskite NdAlO$_3$, $\alpha$ -- corundum ($\alpha$-Al$_2$O$_3$), $\beta$ -- $\beta$-alumina type Al$_{11}$NdO$_{18}$. On the right ordinate the $T$ and phase compositions are given, that should be expected from the experimental phase diagram Fig. \ref{Fig:ternary}.}
\label{fig:PD-calc}
\end{figure}

The resulting section, that is shown in Fig. \ref{fig:PD-calc}, may therefore be regarded as a suitable and self-consistent basis for the discussion of the ``G'' section. This section cannot be quasi-binary as the end member Nd$_3$Al$_5$O$_{12}$ melts incongruenly and can only be understood if the partial system Al$_2$O$_3$--Y$_3$Al$_5$O$_{12}$--NdAlO$_3$ of the concentration triangle Fig. \ref{Fig:ternary} is considered.

In Fig. \ref{fig:PD-calc} the right hand side Nd$_3$Al$_5$O$_{12}$ corresponds to the right end of the ``G'' line in Fig. \ref{Fig:ternary} and should basically bear the same information. However, Fig. \ref{Fig:ternary} is based on a thermodynamic assessment of the whole system Al$_2$O$_3$--Nd$_2$O$_3$ by optimizing $G(T)$ data for all phases, especially $G_\text{ex}(T)$ for the melt. Unfortunately, the quantity and quality of experimental data available for the ternary system Al$_2$O$_3$--Y$_3$Al$_5$O$_{12}$--NdAlO$_3$ is not sufficient to perform such assessment here too, and ideal mixing ($G_\text{ex}=0$) was assumed for the melt as well as for the garnet phase. Irrespectively of these simplifications, the correspondence between Figs. \ref{fig:PD-calc} and \ref{Fig:ternary} is good: For the composition Nd$_3$Al$_5$O$_{12}$ one finds the garnet phase stable up to $930^{\:\circ}$C as indicated by ``\textit{(g)}'' near the right ordinate of Fig. \ref{fig:PD-calc}. Upon further heating the garnet decomposes to a mixture of the perovskite NdAlO$_3$ and $\alpha$-Al$_2$O$_3$ (``\textit{(p+$\alpha$)}'' in Fig. \ref{fig:PD-calc}). At $1393^{\:\circ}$C $\alpha$-Al$_2$O$_3$ disappears and the $\beta$-alumina type Al$_{11}$NdO$_{18}$ is formed: ``\textit{(p+$\beta$)}'' in Fig. \ref{fig:PD-calc}. At $T=1713^{\:\circ}$C Fig. \ref{Fig:ternary} proposes disappearence of Al$_{11}$NdO$_{18}$, resulting in a 2-phase region ``\textit{(l+p)}''. Fig. \ref{fig:PD-calc} slightly underestimates the stability of the melt with respect to Al$_{11}$NdO$_{18}$, hence a ``\textit{(l+p+$\beta$)}'' phase field is found here below ``\textit{(l+p)}''. Fortunately, this minor deviation contributes only to solid/liquid equilibria and no substantial discrepancy in the subsolidus equilibria is evident that are in the focus of the present work.

The (Y$_{1-x}$Nd$_x$)$_3$Al$_5$O$_{12}$ single phase field ``\textit{g}'' extends in Fig. \ref{fig:PD-calc} from $0\leq x\leq0.83$, but is in equilibrium with a liquid phase only for $x\leq0.12$. Following this figure, Nd:YAG with doping levels exceeding $\approx80$\% should be stable on the ``G'' section only with additions of $\alpha$-Al$_2$O$_3$. For $x>0.12$ (Y$_{1-x}$Nd$_x$)$_3$Al$_5$O$_{12}$ can only be obtained by solid and chemical (sol-gel) reactions. The ternary eutectic of the partial system Al$_2$O$_3$--Y$_3$Al$_5$O$_{12}$--NdAlO$_3$ is calculated at $T_\text{eut,t}\approx 1750^{\:\circ}$C. Only this ternary eutectic can be observed at constant $T$ for different compositions $x\gtrsim0.28$. For $0.12\leq x\leq0.28$ the phase boundary ``\textit{g+p}''/``\textit{l+g+p}'' is the projection of a binary eutectic groove with $T_\text{eut,b}\approx 1800\ldots1750^{\:\circ}$C depending on $x$. In DTA heating curves, eutectics are usually marked by sharp endothermal peaks. In the present study $T_\text{eut}^{(25\%)}=1703^{\:\circ}$C and $T_\text{eut}^{(30\%)}=1700^{\:\circ}$C were measured with good reproducibility. The experimental values are by $\approx50$~K below the theoretical values, but the slightly higher $T_\text{eut}$ for the 25\% sample indicates, that $x=0.25$ is already situated on the binary eutectic groove where $T_\text{eut,b}$ depends on $x$. For large Nd concentrations (depending on $T$ for $x>0.12\ldots0.83$) the garnet phase is only stable in equilibrium with perovskite, $\alpha$-Al$_2$O$_3$ (corundum), or Al$_{11}$NdO$_{18}$ ($\beta$).

The subsolidus equilibria depicted in Fig. \ref{fig:PD-calc} can explain the different DTA curves for the first and for subsequent heating runs that are shown in Fig. \ref{fig:DTA}: If the (Y$_{1-x}$Nd$_x$)$_3$Al$_5$O$_{12}$ samples with e.g. $x=0.30$ are heated for the first time, the system starts in the ``\textit{g}'' phase field and crosses subsequently the ``\textit{g+p}'', ``\textit{g+p+}$\beta$'', ``\textit{l+g+p}'',``\textit{l+g}'', and finally the ``\textit{l}'' fields. The first of the transformations ``\textit{g}'' $\rightarrow$ ``\textit{g+p}'' $\rightarrow$ ``\textit{g+p+}$\beta$'' are only transformations between solid phases that are diffusion limited and lead to broad DTA peaks as the reaction is smeared out over a broad $T$ range. If the melt of this composition $x=0.30$ is cooled, the garnet phase crystallizes first with enrichment of Nd in the melt, as $k^\text{Nd}\approx0.15$ \cite{Nishimura75}. From Fig. \ref{fig:PD-calc} one can estimate $k^\text{Nd}=x_\text{sol}/x_\text{liq}\approx0.12$ for small $x$. The remaining melt composition approaches the eutectic point and crystallizes then on the path ``\textit{l+g+p}'' $\rightarrow$ ``\textit{g+p+}$\beta$'' $\rightarrow$ ``\textit{g+p}''. Indeed, the perovskite phase was found in Nd:YAG fibres that were grown in this composition range \cite{Lipinska06}. However, the cooled solid is in a metastable nonequilibrium state where Y-rich garnet occurs together with Nd-rich garnet and with NdAlO$_3$ or even Al$_{11}$NdO$_{18}$. If this metastable sample is heated again it tends to reach equilibrium by a thermally activated exothermal solid state reaction that is responsible for the exothermal DTA peaks.

\section{Summary}

The assumption of the existence of neodymium aluminum garnet allows to compile a self-consistent YAG--NdAG section of the Y$_2$O$_3$--Nd$_2$O$_3$--Al$_2$O$_3$ phase diagram that explains the measured DTA curves for heavily Nd-doped YAG powders. Accordingly, the solubility of Nd$^{3+}$ in YAG ranges up to 80~at\%, i.e.\ is much higher than supposed so far. However, solubility decreases strongly with increasing temperature. Therefore, and due to distinct segregation, the Nd$^{3+}$ concentration in crystals grown from the melt is limited to $\lesssim\,12$~at\%.

\ack{

This work was partly supported by Polish Ministry of Education and Science under the research project number 3 11TB 004 30.}


\end{document}